# FlashMap: A Flash Optimized Key-Value Store

Zonglin Guo
Dept. of Computer Science
University of California, Irvine

Tony Givargis
Dept. of Computer Science
University of California, Irvine

**Abstract**

Key-value stores are a fundamental class of NoSQL databases that offer a simple yet powerful model for data storage and retrieval, representing information as pairs of unique keys and associated values. Their minimal structure enables exceptionally fast access times, scalability, and flexibility in storing diverse data types, making them ideal for high-performance applications such as caching, session management, and distributed systems. As modern computing increasingly demands responsiveness and scalability, key-value stores have become a critical component of the data infrastructure in both industry and research contexts. In this work, we present FlashMap, a high-performance key-value store optimized for Flash-based solid-state drives (SSDs). Experiments show that FlashMap achieves outstanding throughput, averaging 19.8 million inserts and 23.8 million random lookups per second with a 100-byte payload, all on a single data center-grade server.

## 1. Background

A key-value store is a type of database that stores data in a simple, associative array format, where each piece of data (the value) is associated with a unique key. This structure allows for fast access and retrieval of data, typically using the key as a reference.

Key-value stores are widely used in various industries due to their simplicity, scalability, and performance characteristics. They are particularly well-suited for scenarios where quick lookups, flexibility, and easy scaling are essential. Let us outline a number of use cases for key-value stores.

**Session Storage:** Key-value stores are commonly used for session storage in web applications. Sessions involve storing user-specific data (such as authentication tokens, preferences, or shopping cart contents) that needs to be accessed quickly across multiple requests [1].

**Caching:** Key-value stores are heavily used for caching data to reduce the load on primary databases and improve application performance. Caching involves temporarily storing frequently accessed data in a faster, more accessible store (like RAM), and key-value stores are an excellent fit for this [Memcached].

**Configuration Management:** Key-value stores are also used for managing configuration settings in modern software systems. Applications often have a large number of configuration parameters that are stored as key-value pairs [2].

**Real-Time Analytics:** Key-value stores are useful in real-time analytics scenarios where large volumes of data need to be ingested, processed, and queried quickly. They allow applications to store metrics, logs, or other time-series data in a simple key-value format [3].

**Distributed and Scalable Systems:** Key-value stores are naturally well-suited for distributed systems and cloud-native applications. They allow for high scalability because adding more nodes to a cluster can expand both storage and computational power seamlessly [4].

**Event Sourcing:** Key-value stores are often used in event sourcing, an architectural pattern where changes in state are stored as a sequence of immutable events rather than the current state. These events can be replayed to recreate the system's state [5].

**Mobile & Edge Computing:** Key-value stores are also used in mobile and edge computing environments, where devices might need to store local data temporarily before syncing with a central server. Due to their simplicity, key-value stores are ideal for low-latency data storage on resource-constrained devices [6].

Although key-value stores [7] are simple in terms of usage and operational semantics, designing a high-performance implementation involves complex trade-offs that balance the performance and efficiency of core operations. These challenges are further intensified when targeting Flash storage systems, which introduce unique constraints and opportunities. In this work, we present the design and architecture of FlashMap, a key-value store optimized for Flash. We motivate our main design decisions and demonstrate how they contribute to achieving high throughput and low latencies when deployed on Flash-based storage.

The remainder of this paper is organized as follows. In Section 2, we describe the core key-value store operations and associated semantics. In Section 3, we outline the architecture of FlashMap. In Section 4, we describe our benchmarking and performance results. In Section 5, we provide a summary and conclude the paper.

## 2. Key-Value Operations & Semantics

Without loss of generality, we assume that a key *K* in a key-value pair is a non-empty blob of bytes (i.e., its size is ≥ 1), while a value *V* is a blob of bytes of size ≥ 0. A logical key-value store is defined as a named collection containing zero or more key-value pairs. We categorize key-value operations into three groups: resource management operations, point operations, and transaction operations.

**Resource management operations** include *open()* and *close()*.

The *open(storage, name)* function takes a storage specification as an argument, which may refer to a raw block device, a file, or a directory within a file system. This storage medium serves as the persistent backing store for all data, settings, and metadata. The name argument specifies the logical name of the key-value store. On successful execution, *open()* returns a handle that must be used for all subsequent key-value operations.

The *close(handle)* function takes as input a handle previously returned by *open()*. It closes the associated key-value store and releases all allocated resources. Additionally, any key-value pairs held in a cached state are flushed to persistent storage upon successful completion of *close()*.

**Point operations** include the mutation functions: *insert()*, *update()*, *replace()*, *delete()* and the read functions: *lookup()*, *next()*, *prev()*. A point operation executes atomically, transitioning the store from one consistent state to another (i.e., if mutating). These functions take as an argument a *handle* previously obtained via a call to *open()*. For brevity, the *handle* is omitted from the argument lists in the following descriptions, but it should be assumed to be implicitly required.

The functions *insert(K, V)*, *update(K, V)*, and *replace(K, V)* all attempt to store the key-value pair *(K, V)* in the logical key-value store. However, each function has distinct semantics:

- *insert(K, V)* unconditionally adds the *(K, V)* pair to the store, replacing any existing pair with the same key *K* if one exists.
- *update(K, V)* replaces the value of an existing key *K* with *V* if the key is already present; otherwise, it behaves like *insert()* and adds the new *(K, V)* pair.
- *replace(K, V)* updates the value for key *K* only if the key already exists in the store; if no such key is found, the operation fails without modifying the store.

The *delete(K)* function removes the key-value pair associated with key *K* from the store, if such a pair exists.

The *lookup(K)* function returns the value associated with key *K*, if it exists; otherwise, the operation fails. The *next(K)* function returns the *(K, V)* pair corresponding to the lexicographical successor of key *K*, while the *prev(K)* function returns the *(K, V)* pair corresponding to the lexicographical predecessor of key *K*.

Note that *prev()* and *next()* are neighborhood search functions and do not require the search key *K* to exist in the store. If *K* is omitted in a call to *prev()*, the function returns the lexicographically smallest key-value pair. Similarly, if *K* is omitted in a call to *next()*, the lexicographically largest key-value pair is returned.

**Transaction operations** include *transact()*, *commit()*, and *discard()*.

The *transact(parent-handle)* function returns an opaque child handle that represents a logical store containing a snapshot of the key-value pairs present in the parent store at the time the function was called. This child store can be used independently for a sequence of point operations, as described earlier. Conceptually, invoking *transact()* creates a fork in the key-value store, with the parent and child stores evolving separately from that point onward.

At a later time, the child store may be discarded using *discard(child-handle)*, which releases all associated resources. Alternatively, the child store can be merged back into the parent using the *commit()* function. The merge follows these rules:

- If a key-value pair *(K, V)* exists in both the parent and child stores and the values are identical, it remains unchanged.
- If the same key *K* exists in both stores but with different values, the value from the child store overrides that of the parent.
- If a *(K, V)* pair exists only in the child, it is added to the merged store.
- If a *(K, V)* pair exists only in the parent, it is removed in the merged result.

After a successful merge, the merged store becomes the new parent, and the child store is discarded. Both *transact()* and *commit()* are atomic operations, ensuring consistency of the store throughout the process [9].

## 3. FlashMap Architecture

Our proposed FlashMap architecture is designed by leveraging key characteristics of flash memory-based SSDs [8]. Specifically, we assume that the underlying block-oriented storage device favors large sequential writes (e.g., 64 KB) and small random reads (e.g., 4KB). We also assume the device supports a degree of hardware parallelism, with 4, 8, 16, 32, or 64 channels potentially operating in parallel under ideal conditions. Additionally, the storage system performs periodic garbage collection, which, if not accounted for, can negatively impact read/write latency at the application level. Finally, we assume that the device can more efficiently manage its internal resources when stale data is trimmed by the application in large, contiguous chunks.

### 3.1 The Append-Only Strand

Building on this design, the FlashMap architecture introduces the concept of *N + M* strands, where each strand is an append-only, log-structured array of size *S*. The size *S* is typically set to the total capacity of the storage device divided by *N + M*. Each strand supports three operations: *strand.append(K, V)* and *strand.lookup(A)*, and *strand.scan()*. The *strand.append()* function serializes the *(K, V)* pair and appends it to the strand's tail, returning a unique address *A*. The *strand.read()* function retrieves and deserializes the *(K, V)* pair stored at address *A*. Serialized pairs include metadata (i.e., meta structure) to fully describe the lengths of the key and value, along with additional bookkeeping state information. A *strand.scan()* operation on a strand, sequentially recovers each pair in the order it was appended. This function can be used to enumerate the set of addresses corresponding to the key-value pairs stored in the strand.

The following figure illustrates the internal state of a sample strand at a certain point during its lifetime.

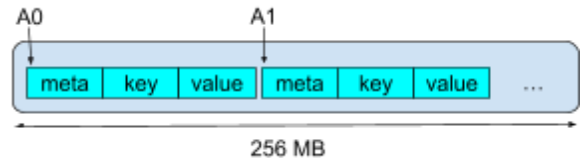

In this figure, *A0* is the address of the first key-value pair and equals 0. The meta structure of a key-value pair is a fixed length region that contains at least the following essential fields.

```
meta {
    link;
    key_len;
    val_len;
}
```

As their names suggest, *key_len* and *val_len* store the lengths of the key and value bytes that follow. The *link* field is used to chain multiple *(K, V)* pair updates that share the same key, as outlined later in this section. In practice, additional fields may be included to improve the robustness of the key-value store against data corruption. For example, a sentinel value can be placed at the beginning of the meta structure to assist in scanning and recovering key-value pairs. Similarly, a signature or checksum may be added to validate the integrity of the entire key-value pair. These fields are omitted here for brevity. It is also important to keep the meta structure compact. In particular, the bit-width of each field should be carefully chosen to support the maximum expected key-value pair size while minimizing overhead.

At a fundamental level, implementing *strand.append()*, *strand.read()*, and *strand.scan()* is relatively straightforward. In practice, however, these operations must be both atomic and performant under high thread concurrency. To ensure optimal I/O performance, *strand.append()* should incorporate buffering, while *strand.lookup()* must leverage object caching.

Inserting a *(K, V)* pair into a strand is as simple as calling *strand.append()*. Deletion is performed by appending a sentinel value, for example, *(K, -1)* where *meta.val_len = -1* indicates a tombstone. Updating a *(K, V)* pair involves appending the new version and setting its *meta.link* to point to the previous version.

With these mutation rules, a strand becomes an append-only structure capable of fully tracking valid, deleted, and updated key-value pairs. The only additional requirement is an index that maps each key *K* to its corresponding address *A*.

### 3.2 The Index

FlashMap can use any data structure capable of efficiently supporting the following operations:

- *index.update(K, A)* associates the key *K* with the address *A*. Ideally, this operation should have a time complexity of O(1) or O(log N).
- *index.lookup(K)* returns the address associated with key *K*, or an error if the key is not present. This operation should also target O(1) or O(log N) time complexity.
- *index.next(K)* and *index.prev(K)* return the address associated with the lexicographical successor or predecessor of *K*, respectively, again with a desired time complexity of O(1) or O(log N).

The index is designed to reside entirely in memory and must therefore maintain a low memory footprint. Suitable data structures include binary or ternary search trees, succinct data structures, and hash tables with sorted indices.

FlashMap's default built-in index uses a succinct, tree-like structure that achieves O(log N) time complexity for all supported operations, with an average memory usage of just 12 bytes per key.

### 3.3 Top-Level Architecture

Given *N* strands and an index, top-level key-value point operations follow a two-step process. First, the key *K* is hash-mapped to one of the N strands, say *Si*. Then, the index and strand *Si* functions are used to complete the operation. Pseudocode for several of these operations is provided below.

```
update(K, V) {
   i = hash(K) % N
   A = index.lookup(K)
   if (INVALID(A)) { // fresh insert
      A = strand[i].append(K, V, meta.A=-1)
   } else { // update existing
      A = strand[i].append(K, V, meta.A=A)
   }
   index.update(K, A);
}

lookup(K) {
   i = hash(K) % N
   A = index.lookup(K)
   if (INVALID(A)) {
      return NOT_FOUND
   }
   (meta, K, V) = strand[i].lookup(A)
   return V
}
```

It's worth noting that upon startup, each strand can be scanned to reconstruct the index. Additionally, the index can be serialized and persisted during the *close()* operation, and later deserialized and restored during the *open()* operation.

It's also worth noting that, as presented, strands will eventually reach their maximum capacity. To manage this, FlashMap employs a garbage collection strategy.

Recall that when the concept of a strand was introduced, we specified a total of *N + M* strands. Of these, *N* strands are used during normal operation, while the remaining *M* are reserved for garbage collection.

When garbage collection is triggered, a subset of the *N* active strands is scanned to identify all valid key-value pairs. These are then transferred to one of the unused *M* strands. Once the transfer is complete, strand indices are swapped under a brief lock, and the original strands are trimmed. Thanks to the append-only nature of strands, garbage collection can proceed in parallel with normal operations, requiring synchronization only at the final stage of the swap. Pseudocode for garbage collection is provided below.

```
gc(i, j) { // i: old strand, j: new strand
   for (K, V) in strand[i].start_enum() {
      A = index.lookup(K)
      if (VALID(A)) {
         strand[j].update(K, V)
   strand[i].lock()
    for (K, V) in strand[i].continue_enum()
{
      A = index.lookup(K)
      if (VALID(A)) {
         strand[j].update(K, V)
   strand[i].swap(j)
   strand[j].unlock()
   storage.TRIM(strand[i].region())
}
```

This process can be repeated until all *N* strands have been cleaned. FlashMap relies on a real-time control system to anticipate when the active strands are likely to fill up, invoking garbage collection just in time to prevent exhaustion. The *M* garbage collection threads run in the background, with minimal priority, just enough to ensure timely cleanup without impacting foreground performance.

By utilizing the storage device's TRIM [10] command, we ensure that device-level garbage collection is invoked primarily in response to data store demands and, more importantly, under optimal conditions.

Finally, transactions are implemented by creating a temporary data store backed by an in-memory storage emulation layer, which serves as the child state. Upon a call to *commit()*, the primary data store is locked as necessary during the merge process.

## 4. Experiments

To evaluate the performance characteristics of FlashMap, we used an i8g.4xlarge AWS instance, equipped with 16 virtual CPUs, 128 GB of RAM, and a 3,750 GB high-performance SSD. The i8g instance family is powered by Arm-based AWS Graviton4 processors and third-generation AWS Nitro SSDs [11], offering the highest I/O performance available on Amazon EC2 for storage-intensive workloads [12].

We established baseline performance metrics for our system as configured, measuring 874K IOPS for 4 KB random reads and 4.3 GB/s throughput for large sequential writes. We report only these two metrics, as they correspond to the specific I/O patterns used by FlashMap, namely, random small reads and large sequential writes.

As a preparation step, we installed Ubuntu 18.04 and performed 8 TB of random data writes to the SSD, equivalent to more than two full-capacity writes, to ensure the SSD reached a steady-state condition.

In the next step, our test began by populating an empty key-value store with 1 billion key-value pairs, each 100 bytes in size. We then read the entire key-value store in order (i.e., using the same lookup order as the original inserts) followed by a second read with lookups performed in random order. Finally, all key-value pairs were deleted. This procedure was repeated using varying numbers of worker threads to assess performance under different levels of concurrency. This initial test provided a baseline for evaluating FlashMap's performance. Our throughput numbers are summarized in the table below. All values are in millions of operations per second.

| Threads | Insert | Seq. Lookup | Rnd. Lookup | Delete |
|---|---|---|---|---|
| 1 | 1.3 | 1.8 | 1.6 | 1.5 |
| 2 | 2.5 | 3.4 | 3.0 | 2.7 |
| 4 | 4.8 | 6.6 | 6.0 | 5.2 |
| 8 | 10.1 | 13.9 | 12.5 | 11.2 |
| 16 | 19.8 | 26.5 | 23.8 | 21.2 |
| 32 | 18.3 | 24.3 | 21.0 | 20.1 |

In the next test, we measured operation latency under increasing performance demands. Specifically, we recorded the maximum latency at which 95%, 99%, and 99.9% of all operations completed. The I/O workload consisted of 100-byte objects with a mix of 20% random updates and 80% random lookups. The resulting latency metrics are summarized in the table below, with all values reported in microseconds.

| Threads | 95% | 99% | 99.9% |
|---|---|---|---|
| 1 | 0.65 | 2.82 | 8.07 |
| 2 | 0.72 | 3.07 | 8.90 |
| 4 | 0.94 | 3.10 | 11.73 |
| 8 | 1.37 | 2.97 | 16.86 |
| 16 | 1.70 | 3.03 | 20.99 |
| 32 | 2.56 | 6.91 | 31.72 |

It is important to note that in all our tests, FlashMap employed *write buffering* and *read caching*. Throughout the tests, we monitored SSD throughput and confirmed that FlashMap was reading from and writing to the SSD at its maximum physical throughput. The bulk of the performance was driven through the caching subsystem. This benefit will diminish for large working datasets. FlashMap uses 32 strands plus 1 additional strand for garbage collection. Each strand utilized a 32 MB read buffer and a 1 GB key-value pair read cache.

The tests above demonstrate the efficiency of FlashMap's internal data structures (such as indexing, write buffering, and read caching) particularly under high thread concurrency, where synchronization and mutual exclusion are required. For instance, with 16 threads, FlashMap can perform 23.8 million random lookups of 100-byte key-value pairs. As expected, increasing the key-value payload size shifts the performance bottleneck to the SSD's physical read and write throughput and garbage collection efficiency. In workloads involving large key-value pair reads and writes, the specific design of the key-value store becomes less critical. However, as the size of key-value pairs decreases, architectural and implementation details have a much greater impact on overall performance. The figure below shows the throughput of FlashMap using 16 threads, with a workload consisting of 20% random updates and 80% random lookups, as a function of key-value size [13]. To eliminate the impact of read caching and highlight the worst-case performance constrained by the SSD's physical throughput limits, the total key space is set to 500 GB.

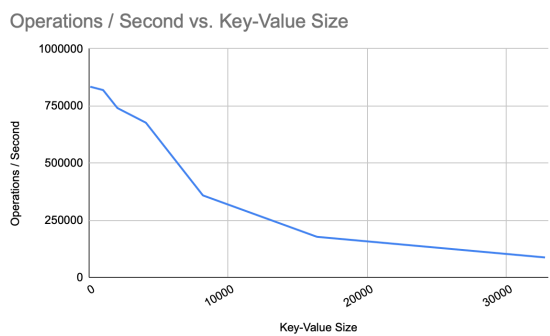

FlashMap can fully utilize the physical SSD storage when the working set exceeds the caching capacity. Conversely, it delivers a high-throughput, low-latency key-value service backed by persistent storage, even under heavy application load.

## 5. Conclusion

As demand for low-latency, high-throughput data systems grows, key-value stores have become critical in both industry and research. Key-value stores continue to play a pivotal role in today's system designs. In this paper, we presented FlashMap, a high-performance key-value store optimized for Flash-based SSDs. FlashMap achieves up to 19.8 million inserts and 23.8 million random lookups per second (100-byte payloads) on a single data center-class server.

## 6. Citations

[1] Sanfilippo, Salvatore. Redis. 2009. https://redis.io. Accessed 4 Oct. 2025.

[2] HashiCorp. Consul. HashiCorp, 2014, https://www.consul.io. Accessed 4 Oct. 2025.

[3] The Apache Software Foundation. Apache Cassandra. Version 5.0, The Apache Software Foundation, 2008, https://cassandra.apache.org. Accessed 4 Oct. 2025.

[4] Amazon Web Services, Inc. Amazon DynamoDB. Amazon, 2012, https://aws.amazon.com/dynamodb/. Accessed 4 Oct. 2025.

[5] Event Store Ltd. EventStoreDB. Event Store Ltd., 2012, https://eventstore.com. Accessed 4 Oct. 2025.

[6] Hipp, D. Richard. SQLite. SQLite Consortium, 2000, https://sqlite.org. Accessed 4 Oct. 2025.

[7] DeCandia, Giuseppe, et al. "Dynamo: Amazon's Highly Available Key-value Store." ACM SIGOPS Operating Systems Review, vol. 41, no. 6, 2007, pp. 205-220.

[8] Takeuchi, Ken, Teruyoshi Hatanaka, and Shuhei Tanakamaru. "Highly Reliable, High Speed and Low Power NAND Flash Memory-Based Solid State Drives (SSDs)." IEICE Electronics Express, vol. 9, no. 8, 2012, pp. 779-794.

[9] Gray, J., & Reuter, A. (1992). Transaction processing: Concepts and techniques. Morgan Kaufmann.

[10] Frankie, T. C. (2012). Model and Analysis of Trim Commands in Solid State Drives (Doctoral dissertation, University of California, San Diego).

[11] Amazon Web Services, Inc. "AWS Nitro SSD – High Performance Storage for Your I/O-Intensive Applications." AWS News Blog, 13 Mar. 2024.

[12] Amazon Web Services, Inc. https://aws.amazon.com.

[13] Pelley, S., & Tufte, J. (2015). Scalable Key-Value Store Designs: A Case Study. Proceedings of the 2015 International Conference on Cloud Computing and Big Data, 42-49.